\documentstyle[multicol,aps,psfig]{revtex}
\begin{document}

\draft

\title{Arbitrarily Accurate Eigenvalues for One-Dimensional Polynomial 
Potentials}

\author{Y. Meurice \\ 
{\it Department of Physics and Astronomy, The University of Iowa, 
Iowa City, Iowa 52242, USA}}

\maketitle
\begin{abstract}
We show that the Riccati form of the one-dimensional Schr\"odinger 
equation can be reformulated in
terms of two linear equations depending on an arbitrary function $G$.
When $G$ and the potential (as for anharmonic oscillators) 
are polynomials
the solutions of these two equations are entire 
functions ($L$ and $K$) and  
the zeroes of $K$ are identical to those of the wave function.
Requiring such a zero at a large but finite value of the argument 
yields the low energy eigenstates with exponentially small errors.
Approximate formulas for these errors are provided.
We explain how to chose $G$ in order 
improve dramatically the numerical treatment.
The method yields many significant 
digits with modest computer means. We discuss the extension of this 
method in the case of several variables.

\end{abstract}
\pacs{PACS: 03.65.Ge, 03.65.-w, 02.30.Em, 02.30.Mv, 10.10.St, 33.20.-t}

\begin{multicols}{2}\global\columnwidth20.5pc
\multicolsep=8pt plus 4pt minus 3pt
 
\section{Introduction}
Quantum anharmonic oscillators appear 
in a wide variety of problems in molecular, 
nuclear or condensed matter physics. 
Typically,
anharmonic terms appear in expansions about a minimum of a potential, when 
ones tries to incorporate the non-linear features of the 
forces responsible for this 
equilibrium. 
The most celebrated example is the quartic anharmonic oscillator 
\cite{bender69}
where a $\lambda x^4$ term is added to the usual harmonic Hamiltonian.
Introducing bilinear couplings among a set of such oscillators leads to
a rich spectrum, for instance, multiphonon bound states in one-dimensional
lattice models \cite{wang96}.
More generally, one can think about the $\lambda \phi^4$ (or 
higher powers of $\phi$) field theories in
various dimensions as systems of coupled anharmonic oscillators.

Anharmonic terms can be treated perturbatively and the 
perturbative series can be represented by Feynman diagrams. Unfortunately,
the coefficients of the series\cite{bender69,leguillou90} 
have a factorial growth 
and the numerical values obtained from the 
truncated series have an
accuracy which is subject to limitations. At fixed coupling, there is an 
order at which an optimal accuracy is reached. At fixed order, there is 
a value of the coupling beyond which the numerical values are meaningless
even as an order of magnitude. In the case of the
single-well quartic potential, Pad\'e approximants can be used for the
series or its Borel transform. Rigorous proofs of convergence can 
be established in particular cases \cite{loeffel69}. 
Unfortunately, such a method does not apply to the case of the double-well 
potential\cite{brezin77} 
where instanton effects \cite{coleman,zj} need to be taken into 
account. It should also be noted that even when Pad\'e approximants converge,
the convergence rate may be slow. Strong coupling
expansions \cite{weninger96} or variational interpolations \cite{kleinert}
sometimes provide more accurate results. 

The above discussion shows that 
finding an expansion which can be used {\it indiscriminately} for 
most quantum mechanical problems with polynomial potentials
remains a challenging problem.
Alternatively, one can use numerical methods. Variational methods
are often used to obtain upper and lower bounds on energy levels 
\cite{bazley,payne}. These methods are based on rigorous inequalities
and are considered superior to methods based on numerical integration 
\cite{payne}. However, the difference between the bounds widens rapidly
with the anharmonic coupling and the energy level. 
Methods based on series expansions in the position variable
\cite{biswas,kbeck81,fernandez,bacus} appear to produce more significant digits
more easily. However, our understanding of 
the convergence and numerical stability of these methods seems to be limited
to empirical observations. The methods  based on series expansions fall into 
two categories: methods based on the 
evaluations of determinants \cite{biswas,fernandez} and methods based on
boundary conditions at large but finite values of the position 
\cite{kbeck81,bacus}. The main goal of this article is to provide 
a systematic discussion of the errors associated with this second category
of methods and to show how to make these errors 
arbitrarily small in the most 
efficient way. With the exception of Section \ref{sec:multi}, we 
only consider one-dimensional problems. We discuss two types of errors.
First, the numerical errors made in calculating the energy which makes the 
wave function vanish at some large value of the position $x_{max}$.
Second, the intrinsic error due to the finiteness of $x_{max}$.

The basic elements the numerical method used hereafter 
were sketched in Ref.\cite{bacus}
and applied to the quartic anharmonic oscillator. 
We wrote the logarithmic
derivative of the wave function which appears in the Riccati equation 
as $L/K$ and showed that these functions were entire.
The values of 
the first ten eigenvalues with 30 significant digits 
provided for a particular coupling 
have been used to test new theoretical methods\cite{antonsen}. 
Two issues were left open in this formulation: 
first, the basic equations 
had an interesting invariance which was not undestood but could 
be used to improve the numerical efficiency; second, 
the use of the method for
parity non-invariant potentials appeared to be unduly complicated \cite{oktay}.

In Section \ref{sec:basic},
we present  a new formulation where these two
issues are settled. 
The basic equations presented 
depend on an arbitrary {\it function}
denoted $G(x)$.
This freedom
can be interpreted as a local gauge invariance associated with 
the fact that only $L/K$ is physical. 
The wave function is invariant under 
these local transformations. 
In section \ref{sec:sol}, we show 
how to construct power series for $L$ and $K$. 
The complications in the case of parity non-invariant potentials (such as 
asymmetric double-wells) are 
minimal. When the potential and the gauge function are polynomials, 
these series define {\it entire} function. 
In other words, it is always possible to 
construct arbitrarily accurate solutions of the  Schr\"odinger equation
for arbitrary $E$ within a given range of the position variable, 
by calculating enough terms in the expansions of $L$ and $K$. This allows
us to reproduce the asymptotic behavior of the wave function and determine 
the energy eigenvalues. In section \ref{sec:sens}, we use the global properties
of the flows of the Riccati equation to recall of some basic results related
to the WKB approximation and the Sturm-Liouville theorem. 
We explain how bifurcations in the asymptotic behavior of the functions 
$K$ and $L$ can be exploited to determine the eigenvalues.

It should be noted that the importance of reproducing the proper 
asymptotic behavior has been emphasized in variational approaches 
\cite{turbiner84}.
It should also be noted that Pad\'e approximants have been used in
conjunction with the Riccati equation in Ref. \cite{fernandez}, where
the quantization condition used 
was that the approximants give one 
additional coefficient in the Taylor expansion. 
This procedure depends only on the coefficients of the expansions used and
there is no reference to any particular value of $x$
(as our $x_{max}$). Consequently, there is no 
obvious connection between the two approaches.

In the next two sections, 
we show how to turn the gauge invariance to our advantage. 
In Section \ref{sec:bif}, the quantitative aspects of 
the bifurcation are discussed with an exponential parametrization 
similar to the one 
used to determine Lyapounov exponents in the study of chaotic dynamical 
system. The exponents are $G$-dependent. 
We provide an approximate way to determine the exponents and 
the energy resolution.
We explain how our freedom in chosing $G$ can be used to 
make the bifurcation more violent and improve the energy 
resolution. However, the choice of $G$ also affects 
the convergence of $L$ and $K$ and consequently the numerical
accuracy of the solution of the Schr\"odinger equation.
In Section \ref{sec:opt}, 
we show in a particular example that for an expansion of $L$ and $K$ at
a given order, a judicious choice of gauge can improve tremendously 
the numerical accuracy of an energy level. We 
discuss the two principles which allow to make optimal choices of $G$ 
and provide practical methods to determine approximately this optimal choice
for the general case.
We use these methods to explain some empirical results found in \cite{kbeck81}.

In Section \ref{sec:app}, we discuss the 
the error $\delta E$ on the energy levels due to the finiteness of $x_{max}$. 
We propose two approximate formulas valid, respectively, for intermediate 
and large values of $x_{max}$
and compatible in overlapping ranges.
Note that one can reinterpret the condition 
that the wavefunction vanishes at $x_{max}$ as coming from
a slightly different problem where the potential becomes infinite
at $x_{max}$. 
In the path-integral formulation (which can be extended immediately 
to field theory problems), the fact that the potential becomes infinite
at $x_{max}$ means that 
paths with values of $x$ larger than $x_{max}$ are not taken into account.
It has been argued \cite{pernice98,convpert} 
that these configurations
are responsible for the asymptotic behavior of the 
regular perturbative series. In Ref. \cite{convpert}, we showed that 
the perturbative series of several modified problem were convergent.
The error formula sets the accuracy limitations 
of this approach. Some of the methods used in this section could be used 
for quantum field theory problems.

The anharmonic oscillator can be considered as a 
field theory with one time and zero space dimensions. 
It can be used to test
approximate methods such as perturbative expansions or 
semi-classical procedures. An illustrative 
example is given in Ref. \cite{jentschura}
where multi-instanton effects were considered and where the splitting of 
the two lowest levels of a double-well problem were estimated with more 
than hundred digits. In Section \ref{sec:chall}, we show that our method 
can be used to reproduce all these digits. Finally, we discuss the 
generalization of the method to problems with several variables 
in Section \ref{sec:multi}. For these problems, our ability to reduce 
the degree of expansion by using optimal gauge functions may be crucial.

\section{Basic equations and their gauge-invariance}
\label{sec:basic}

We consider a one-dimensional, time-independent 
Schr\"odinger equation $H\Psi=E\Psi$, for an Hamiltonian
\begin{equation}
H={p^2\over{2m}}+\sum_{l=1}^{2l}V_jx^j\ .
\label{eq:ham}
\end{equation}
As is well-known, 
one can reexpress  the wave function in terms of its logarithmic
derivative
\begin{equation}
\Psi(x)\propto{\rm e}^{-{1\over \hbar}\int_{x_0}^{x}dy \phi(y)\ ,}
\label{eq:repa}
\end{equation}
and obtain the Riccati form of the equation:
\begin{equation}
\hbar \phi '=\phi^2+2m(E-V)\ .
\label{eq:ric}
\end{equation}
It is assumed that $m>0$ and that the leading power 
of $V$ is even with a positive
coefficient ($V_{2l}>0$).

Writing $\phi=L/K$, we obtain a solution of Eq. (\ref{eq:ric}) provided
that we solve the system of equations:
\begin{eqnarray}
\label{eq:basic1}
\hbar L'&+&2m(V-E)K+GL=0 \\
\label{eq:basic2} 
\hbar K'&+&L+GK=0
\end{eqnarray}
where $G(x)$ is an unspecified function.
This can be seen by multiplying (\ref{eq:basic1}) by 
$K$, (\ref{eq:basic2}) by $L$ and
eliminating $GKL$ by taking the difference. 
One then obtains the Riccati equation (\ref{eq:ric})
multiplied by $K^2$. Near a zero of $K$, one can check that Eqs. 
(\ref{eq:basic1}-\ref{eq:basic2}) remain valid, 
namely they impose that $\phi$ has a simple pole
with residue $-\hbar$. This allows the wave function to 
become zero and change 
sign as the contour goes around the pole on either side.

Eqs. (\ref{eq:basic1}-\ref{eq:basic2})
 are invariant under the {\it local} transformation
\begin{eqnarray}
\nonumber
L(x)&\rightarrow & Q(x)L(x) \\ 
K(x)&\rightarrow & Q(x)K(x) \\
\nonumber 
G(x)&\rightarrow & G(x)-\hbar Q'(x)/Q(x) \ ,
\label{eq:gt}
\end{eqnarray}
where $Q(x)$ is an arbitrary function.
It is clear that this transformation leaves $\phi$ and the 
wave function unchanged.
If we choose $G=0$ and eliminate $L$ using Eq. 
(\ref{eq:basic2}), we recover the Schr\"odinger equation for $K$.
Starting from this gauge and making an arbitrary transformation, we find
that in general
\begin{equation}
K(x)\propto\Psi(x){\rm e}^{-{1\over \hbar}\int_{x_0}^{x}dy G(y)}
\label{eq:kgen}
\end{equation}
This shows that when $G$ is polynomial, 
$K$ is simply $\Psi$ multiplied
by an entire function {\it with no zeroes} \cite{knopp}. 
This means that the zeroes of 
$K$ and $\Psi$ are identical. In other words, there are no spurious zeroes
when $G$ is polynomial.

By taking the derivative of Eqs. (\ref{eq:basic1}) and (\ref{eq:basic2}) 
and choosing $G(x)$ 
appropriately, 
one can obtain the basic Equations used in \cite{bacus}. 
The explicit form of $G(x)$
is reached by comparing the two sets of equations and integrating 
one of the differences. The two possibilities are compatible. The resulting
integral expression can be worked out easily by the interested reader. 
The only important point is that the $G$ found that way is in general
not polynomial, justifying the spurious zeroes found with the original 
formulation.

\section{Solutions in terms of entire functions}
\label{sec:sol}

The function $G$ can be chosen at our convenience.
For instance, we could impose the condition $K=1$ by 
taking $G=-L$ and recover 
the Riccati equation for $L$. However, the main advantage of 
Eqs. (\ref{eq:basic1}-\ref{eq:basic2}) is that they are linear 
first order differential equations with variables coefficients.
It is well-known \cite{coddington} that if we consider these equations for 
complex $x$, the solutions inherit the domain of analyticity of the
coefficients (provided that this domain is simply connected).
If the coefficients are entire functions, 
there exists a unique entire solution corresponding to a particular 
set of initial values. In the following, we restrict ourselves to 
the case where $V$ and $G$ are 
polynomials.

One can construct the unique solution corresponding to a particular choice
of initial values $L(0)$ and $K(0)$ by series expansions. 
Using $K(x)=\sum_{n=0}^{\infty}K_nx^n$ and similar notations 
for the other functions, one obtains the simple recursion
\begin{eqnarray}
\nonumber
L_{n+1}&=&{-1\over{\hbar (n+1)}}(\sum_{l+p=n}(2mV_lK_p+L_lG_p)
 -2mEK_n)
\\ 
K_{n+1}&=&{-1\over{\hbar (n+1)}}(L_n+\sum_{l+p=n}K_lG_p)
\label{eq:iter}
\end{eqnarray}
Given $L_0$ and $K_0$, these equations allow to determine 
all the other coefficients. 
For potentials which are parity invariant, and 
if $G$ is 
an odd function, $L$ and $K$ can be assigned definite and opposite 
parities. In this case, we can impose the initial conditions $K_0=1$ and 
$L_0=0$ for even wave functions and $K_0=0$ and 
$L_0=1$ for odd wave functions. If the Hamiltonian has no special symmetry,
as for instance in the case of an asymmetric double-well, 
one could leave $L_0$
indeterminate and fix it at the same time as $E$ 
using conditions on the wave function or its 
derivative at two different points. These two conditions translate 
(in good approximation) into 
two polynomial equations in $L_0$ and $E$ and can be solved by Newton's 
method.

The fact that Eqs. (\ref{eq:iter}) determines entire functions provided that
$V$ and $G$ are polynomials can be inferred directly from the fact that 
the coefficients will decrease as $(n!)^{-\kappa }$ for some positive 
power $\kappa$ to be determined and in general 
depending on the choice of $G$. 
As we will explain in more detail in Section \ref{sec:sens}, 
if the leading term in $V$ is $V_{2l}x^{2l}$, 
one expects from Eq. (\ref{eq:ric}) that for $x$ large enough,
\begin{equation}
\phi(x)\simeq\pm \sqrt{2mV_{2l}}x^{l}\ ,
\label{eq:aswkb}
\end{equation}
and asymptotically,
\begin{equation}
\Psi(x)\propto {\rm e}^{- {\pm 1\over{(l+1)\hbar}} \sqrt{2mV_{2l}}x^{l+1}}\ .
\end{equation}
Looking at the general expression for $K$ given in Eq. (\ref{eq:kgen}), one 
sees that $K$ will have the same asymptotic behavior provided that the 
integral of $G$ grows not faster than $x^{l+1}$. If this is the case, 
then $\kappa =1/(l+1)$. This behavior is well observed in empirical series.

Note that if $G$ grows faster than $x^l$, the coefficients decay more slowly
and the procedure seem to be less efficient.
In the following, we will mostly discuss the case $l=2$. If we require that 
$G$ is an odd polynomial growing not faster than $x^2$, this means that 
$G$ is homogeneous of degree 1.

\section{Quantization from global flow properties}
\label{sec:sens}
\begin{figure}
\centerline{\psfig{figure=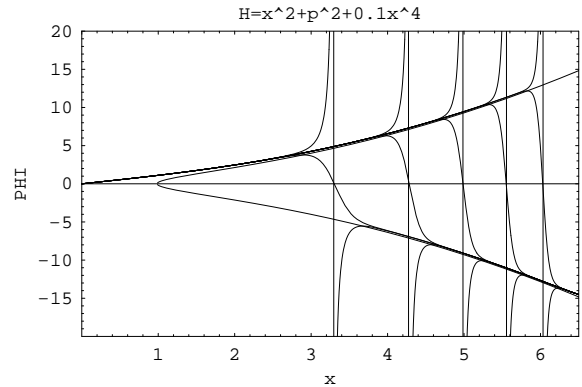,width=3.in}}
\caption{Bifurcations of $\phi(x)$ from the upper part of the WKB curve 
associated with the ground state energy $E_0$ for 
energies $E_0\pm 10^{-5}$, $E_0\pm 10^{-10}$, $E_0\pm 10^{-15}$, 
$E_0\pm 10^{-20}$ and $E_0\pm 10^{-25}$ (from left to right).}
\label{fig:bif1}
\end{figure}

\begin{figure}
\centerline{\psfig{figure=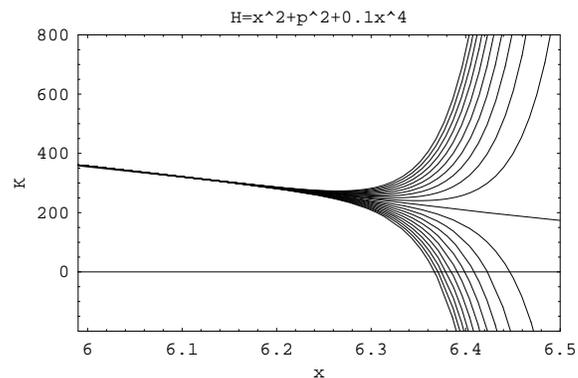,width=3.in}}
\caption{Bifurcations of $K(x)$ from its trajectory for $E=E_0$.
The changes in $E$ are $\pm 10^{-30}$, $\pm 2\times 10^{-30}, \dots$,
$\pm 10^{-29}$ }
\label{fig:kas}
\end{figure}

In this section, we use 
the global properties of the flows associated with the Riccati equation
to rephrase some implications of Sturm-Liouville 
theorem and to justify the asymptotic behavior given in Eq. (\ref{eq:aswkb}).
The main goal of this section is to provide a simple 
and intuitive  picture of the 
bifurcation which occurs when the value of $E$ is varied by a small amount
above or below an energy eigenvalue. The main results of this section 
are summarized 
in Figs. \ref{fig:bif1} and \ref{fig:kas}.

We consider the solutions of Eq. (\ref{eq:ric}) 
obtained by varying $E$ with fixed initial values. It is 
convenient
to introduce an additional parameter $s$ and to rewrite 
the original equation as a 2-dimensional ODE with an $s$-independent 
r.h.s .
\begin{eqnarray}
\label{eq:ode1}
\hbar \dot{\phi}&=&\phi^2+2m(E-V(x))\ \\
\label{eq:ode2}
\dot{x}&=&1\ ,
\end{eqnarray}
where the dot denotes the derivative with respect to $s$.

The flows in the 
$(x,\phi)$ plane have some simple global properties that we now proceed to
describe. We consider a solution (phase curve) with 
initial condition $x=x_0$ and
$\phi=\phi_0$ at $s=0$.
We assume that for these values the r.h.s of Eq. (\ref{eq:ode1}) is $>0$.
It will become clear later that if such a choice is impossible, a 
normalizable wave function cannot be constructed. 
With this assumption, the phase curve starts moving up and right as 
$s$ increases, 
possibly going through simple poles with residues $-h$. This situation persists
unless the r.h.s. of (\ref{eq:ode1}) becomes zero. 
We call the separating curves defined by a 
zero for the r.h.s of Eq. (\ref{eq:ode1}), 
$\phi=\pm \sqrt{2m(V(x)-E})$, ``$WKB$ curves''. After a phase curve crosses
(horizontally) a $WKB$ curve, it moves right and down. If it crosses the 
$WKB$ curve again, we can repeat the discussion as at the beginning. 

At some point, we reach the ``last'' $WKB$ curve 
(i.e., the farthest right). For $x$ large enough, the potential is dominated 
by its largest power and 
the upper (lower) part of this last
WKB curve has a strictly positive (negative) slope. 
For such values of $x$, if a phase curve crosses the WKB curve, 
it will do so 
horizontally and move {\it inside} the region where the r.h.s. of
Eq. (\ref{eq:ode1}) is negative. As $s$ further increases, $\phi$ decreases, 
but the phase curve cannot cross horizontally
the lower part of the WKB curve which has a strictly negative slope. 
In the same region, if $\phi$ 
has a pole, the curve 
reappears below the lower part of the WKB curve and will 
never take positive values again. 

In summary, if in the region described above, 
a phase curve crosses the WKB curve or develops a pole, then it cannot 
develop a pole again. The other logical possibility is that the phase 
curve does none of the above.
It is thus clear that for fixed $E$, we can always find 
a $X$ such that if $x>X$, $\phi(x)$ has no pole. Consequently the two 
terms involving $\phi$ in Eq. (\ref{eq:ric}) cannot grow faster than $2m(E-V)$.
Otherwise, $2m(E-V)$ would become negligible and a pole would be  necessary. 
At least one of these two terms needs to match $2m(E-V)$. Inspection of the
two possibilities leads to Eq. (\ref{eq:aswkb}). Only the positive solution
which follows asymptotically the upper WKB curve 
leads to a normalizable wave function.

If we compare two phase curves with identical initial conditions 
but different $E$, the one with larger $E$ initially lays 
above the other one. If the one with lower $E$ has a first pole at $x_1$, 
then the one with larger $E$ has a first pole at some $x<x_1$. Remembering 
that the poles of $\phi$ produce zeroes of $\Psi$, this  
rephrases the main
idea behind the Sturm-Liouville theorem.
An exact energy eigenstate $E_n$ 
is obtained when the wave function has its last zero
at infinity. When $E$ is fine-tuned to that value, $\phi$ follows
closely the upper branch of the WKB curve. This trajectory in unstable under
small changes in $E$.
If the energy is slightly increased with respect to 
$E_n$, $\phi$ develops a
pole and reappears on the lower part of the WKB curve.  
If the energy is slightly decreased with respect to $E_n$, $\phi$ crosses
the upper part of the WKB curve and reaches the lower part of the WKB curve. 
This is illustrated in 
Fig. \ref{fig:bif1} in the case of the ground state of the quartic single-well
anharmonic oscillator with 
$ m=1/2,\ \hbar=1, V_2=1$ and $V_4=0.1$. All the figures in this section and 
the next two sections have been done with this particular example.
The sensitive dependence on $E$ is also present in the asymptotic
behavior of $K$.
If the energy is slightly increased with respect to $E_n$, $K$ 
reaches zero at a finite value of $x$.
If the energy is slightly decreased with respect to $E_n$, $K$ increases
rapidly. This is illustrated in Fig. \ref{fig:kas} for the same example as 
in Fig. \ref{fig:bif1}.
We now discuss the initial value $\phi_0$. 
For parity invariant potentials, 
one only needs to consider the cases $\phi_0=0$ (even $\Psi$) or $\phi_0=
-\infty$ (odd $\Psi$) at $x_0=0$. For potentials with no reflection symmetry, 
one needs to
insure that the appropriate behavior is reached when $x\rightarrow -\infty$.
This can be implemented in good approximation
by requiring that the wave function 
has also a zero at some large negative value $x_{min}$. 
For potentials with a reflection symmetry about another point $x_1$ than the 
origin, one can impose that the wave function ($K(x_1)=0$) or its derivative 
($L(x_1)=0$) vanish at that point.
In all cases, we have an independent condition 
which allows to determine $\phi_0$.

In summary, for $x_{max}$ large enough, the condition
\begin{equation}
K(x_{max})=0
\end{equation}
provides sharp upper bound on the energy levels. The lower part of 
Fig. \ref{fig:kas} makes clear that as $x_{max}$ increases, sharper
bounds are reached. For potentials that are not parity invariant, an 
additional condition has to be imposed. In all cases, one obtains 
polynomial equations which can be solved for the energy levels given 
the potential or vice-versa using Newton's method. Note also that 
a sharp lower bound can be found by solving $L(x_{max})=0$. The 
fact that in Fig. \ref{fig:kas}, a zero of $K$ at $E_0+\delta$ 
corresponds to a zero of $L$ at  $E_0-\delta$, suggests
that the exact value should be very close to the average of the two bounds.

\section{$G$-dependence of the bifurcation}
\label{sec:bif}

\begin{figure}
\centerline{\psfig{figure=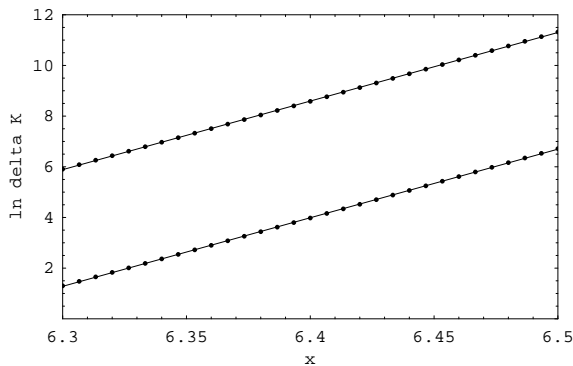,width=3.in}}
\caption{Natural logarithm of $\delta K(x)$ for $E-E_0=10^{-30}$ (lower set 
of point) and $E-E_0=10^{-28}$ (upper set 
of point). Lines are linear fits. }
\label{fig:fitexp}
\end{figure}

\begin{figure}
\centerline{\psfig{figure=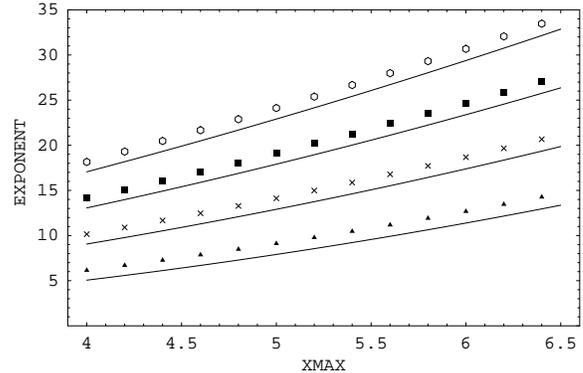,width=3.in}}
\caption{Value of $B$ for various $x$
and for $G=-3x$ (empty hexagons), $G=-2x$ (filled squares) $G=-x$ (crosses) 
and $G=0$ (triangles). The continuous lines have been drawn using 
Eq. (\ref{eq:bap}).}
\label{fig:expdep}
\end{figure}

The strength of the 
bifurcation in $K$ illustrated in Fig. \ref{fig:kas} 
can be approximately characterized by local exponents.
If we consider the departure $\delta K(x)$ from $K(x)$ calculated at some 
exact energy level $E_n$, we expect the approximate behavior:
\begin{equation}
\delta K(x) \simeq C (E-E_n){\rm e}^{xB}\ .
\label{eq:exp}
\end{equation}
In other words ln($|\delta K(x)|$) is linear with a slope $B$ 
independent of the choice of $E$ and an intercept that varies like 
ln$(|E-E_n|)$. This situation is approximately realized in the 
example considered before as shown in Fig. \ref{fig:fitexp}.
We have checked in the same example that the sign of the energy difference 
plays no role. In other words, the same values of $C$ and $B$ can be 
used above and below $E_n$.
The exponent $B$ is not uniform. 
It increases with $x$ and is $G-$dependent
as shown in Fig. \ref{fig:expdep}. The local values of $B$ have been 
calculated by fits in regions of width 0.2 with central value displayed in
the horizontal label of Fig. \ref{fig:expdep}. 
The change in $B$ can be understood as follows. If $E$ is 
changed from $E_n$ to $E_n+\delta E$, then at some point we have a sudden 
transition from the upper to the lower WKB curve and 
asymptotically
\begin{equation}
\delta \Psi (x)\propto \delta E \ 
{\rm e}^{+{1\over{(l+1)\hbar}} \sqrt{2mV_{2l}}x^{l+1}}\ .
\end{equation}
Using Eq. (\ref{eq:kgen}) and expanding about $x_{max}$, we obtain that,  
in good approximation, 
\begin{equation}
B\simeq {1\over \hbar}\bigl(\sqrt{2mV_{2l}}x_{max}^{l}-G(x_{max})\bigr)\ .
\label{eq:bap}
\end{equation}
As shown in Fig. \ref{fig:expdep}, 
this simple expression provides reasonable estimates of $B$. The 
slight underestimation comes in part 
from the fact that Eq. (\ref{eq:bap}) does not take into account the 
harmonic term in $V$.
Eq. (\ref{eq:bap}) shows that we can increase the strength of the 
bifurcation near $x_{max}$ 
by increasing $x_{max}$ or $-G(x_{max})$. 
This allows us to ``resolve'' the energy more accurately. However, at the 
same time our numerical resolution of $K(x_{max})$ is affected 
and we need to take this effect into account. 
This question is treated in 
the next Section. In general, if we can establish that $K(x_{max})$ 
at an energy $E$ very close to $E_n$,
can be calculated with some numerical accuracy $\delta K^{num.}$, we  
have the approximate numerical energy resolution 
\begin{equation}
\delta E^{num.}\propto \delta K^{num.}
{\rm e}^{+{1\over {\hbar}}
\bigl({-1\over{l+1}}\sqrt{2mV_{2l}}x_{max}^{l+1}+\int^{x_{max}}_0
dxG(x)\bigr)}\ .
\label{eq:eresol}
\end{equation}

\section{An optimal choice of $G$}
\label{sec:opt}

In this Section, we show that 
from a numerical point of view, the choice of $G$ is important.
We discuss the question of an optimal choice, first with an example 
and then in general. 
We start with the calculation 
of the ground state in the case
$ m=1/2,\ \hbar=1, V_2=1$ and $V_4=0.1$.
We discuss the estimation of  the ground 
state energy 
using the equation $K(x_{max})=0$ with $x_{max}=6$. The fact that we 
use this finite value for $x_{max}$ creates an error in the 25-th digit 
(see Section \ref{sec:app}).

From the discussion of Section \ref{sec:sol}, it is reasonable to 
limit the discussion to a gauge function of the form
\begin{equation}
G(x)=-ax \ ,
\label{eq:gchoice}
\end{equation}
which using Eq. (\ref{eq:kgen}) implies that 
\begin{equation}
K(x)\propto \Psi(x){\rm e}^{{1\over{2\hbar}}a x^2}\ .
\label{eq:kpart}
\end{equation}
With this restriction, the optimization problem is reduced to the 
determination of $a$. As $a$ increases through 
positive values, the features 
of $\Psi$ are exponentially amplified, making the bifurcation 
displayed in Fig. \ref{fig:kas} more violent.
Ideally, we would like to take $a$ as large as possible. However, if $a$ is 
too large, we may need too many coefficients $K_n$ to get a good 
approximation. If we consider
the problem at a given order, the two requirements of sensitivity and 
accuracy result in a compromise which determines the optimal value of $a$.

\begin{figure}
\centerline{\psfig{figure=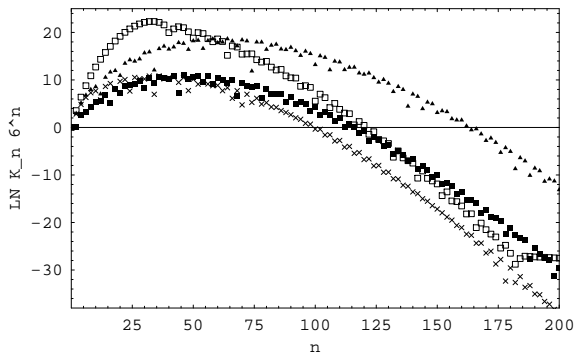,width=3.in}}
\caption{{\rm ln}($|K_n6^n|$) versus $n$, 
for $G=0$ (triangles), $G=-x$ (filled squares), $G=-2x$ (crosses) and 
$G=-3x$ (empty squares). }
\label{fig:k6}
\end{figure}

\begin{figure}
\centerline{\psfig{figure=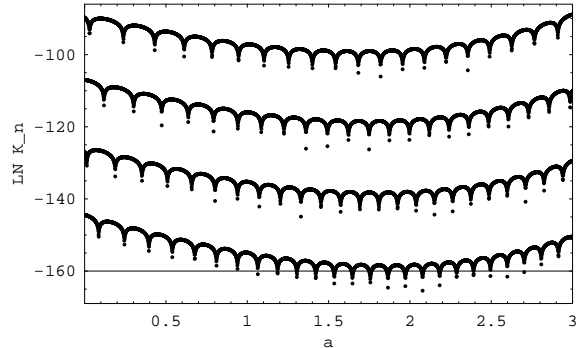,width=3.in}}
\caption{{\rm ln}($|K_{n}|$) versus $a$, for $n$=60 (upper set), 
70 (next set), 80 (next set) and 90 (lower set). }
\label{fig:lncoe}
\end{figure}

As explained in Section \ref{sec:sol}, 
the choice of Eq. (\ref{eq:gchoice}) guarantees a suppression 
of the form $(n!)^{-1\over{3}}$ for the coefficients of $L$ and $K$.
However, the choice of $a$ still affects significantly the behavior
of these coefficients as shown in Fig. \ref{fig:k6}.
The quantity $K_nx_{max}^n$ is relevant to decide at which order we need to 
truncate the series in order to get a good estimate of $K(x_{max})$. 
For instance, if we require knowing 
$K(x_{max})$ with errors of order 1, we 
need about 100 coefficients for $a=2$ but more than 150 for $a=0$. The 
corresponding values
for $a=1$ and 3 fall between these two values, indicating that $a=2$ is close 
to optimal. 
This estimate is confirmed by an analysis of the dependence of $K_n$ on $a$.
Sample values are shown in Fig. \ref{fig:lncoe}. We observe rapid 
oscillations (that we will  not attempt to explain) and
slowly varying amplitudes which have a minimum slightly below 2. 
Note that on the logarithmic scale of Fig. \ref{fig:lncoe}, the 
zeroes of $K_n$ give $-\infty$, however due to the discrete sampling of $a$,
it just generates isolated dots on the graphs. Note also that in Figs. 
\ref{fig:k6} and \ref{fig:lncoe}, the coefficients have been calculated for an 
an accurate value of the ground state energy.
The behavior of the $K_n$ calculated at
value of $E$ sufficiently close to an eigenvalue,
can be understood by using the asymptotic form
\begin{equation}
K(x_{max})\propto 
{\rm e}^{{1\over \hbar}\bigl(-{1\over{l+1}} \sqrt{2mV_{2l}}x_{max}^{l+1}-
\int^{x_{max}}_0dxG(x)\bigr) }\ .
\label{eq:kasy}
\end{equation}
We emphasize that the relative sign between the two terms 
in the exponential is opposite than 
in Eq. (\ref{eq:eresol}), because we are now on the upper WKB curve.
For $a=0$, Eq. (\ref{eq:kasy}) provides a rough estimate of $K_nx_{max}^n$.
Remembering the minus sign in the parametrization of $G$ 
(Eq. (\ref{eq:gchoice})), we see that if 
$a$ is given a small positive value, the argument of the exponential 
in Eq. (\ref{eq:kasy}) decreases and we can obtain comparable accuracy
with less terms in the expansion. Naively, our optimum choice is 
obtained when the two terms in the exponential cancel. 
In the general case, this amounts to having 
\begin{equation}
\sqrt{2mV_{2l}}x_{max}^{l+1}\simeq -(l+1)\int^{x_{max}}_0
dxG(x) \ .
\label{eq:optim}
\end{equation}
For the particular example considered here, this cancellation is 
obtained for $a=(2/3)\sqrt{0.1}x_{max}\simeq 1.27$. It is clear that 
when the two terms cancel, subleading terms neglected in Eq. (\ref{eq:aswkb})
should be taken into account. However, in several examples, we found that
this simple procedure gives results close to what is found empirically.

\begin{figure}
\centerline{\psfig{figure=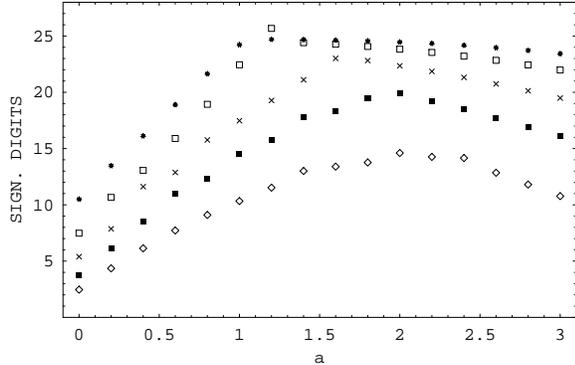,width=3.in}}
\caption{Number of significant digits for $E_0$ versus $a$ using the 
condition $K(6)=0$ with expansions of order 50 (empty diamonds), 75 (filled 
squares),  100 (crosses), 125 (empty squares) and 150 (stars).}
\label{fig:sd}
\end{figure}

We now address the more general question of determining 
the $G$-dependence of the 
number of 
significant digits that can be obtained from the condition 
$K(x_{max})=0$ using an expansion of $K$ 
truncated at a given order. For the example considered before in this section,
we see from Fig. \ref{fig:sd} that, for instance 
for a truncation at order 100, the 
most accurate answer is obtained for $a\simeq 1.6$. 
It is worth noting that for this value of $a$, one gains more than 15 
significant digits compared to the $G=0$ case!  
This figure also indicates,
that as expected, 
the best possible answer (in the present case, 25 significant digits) 
can always be achieved by calculating enough coefficients. 
Using Eq. 
(\ref{eq:eresol}) and Fig. (\ref{fig:lncoe}), we were able to reproduce 
approximately the left part of Fig. \ref{fig:sd} ($0<a<1$). To give a
specific example, at order 100, when one changes $a$ from 0 to 1, 
$\delta K^{num}$ becomes 4 orders of magnitude smaller and the factor 
${\rm e}^{-({a\over{2\hbar}})x_{max}^2}$ improves the resolution by 
almost 8 orders of
magnitude. This approximately accounts for the gain of 11 significant 
digits observed in Fig. \ref{fig:sd}. A detailed understanding of the
figure in the region $1<a<2$ is beyond what can be accomplished using 
the asymptotic form of the wave function. However, the naive estimate 
of Eq. (\ref{eq:optim}) provides a reasonable estimate of the location
of the optimal $a$. 

It should be noted that an ansatz of the form of Eq. (\ref{eq:kpart}) 
with $a=1$ has been used in Ref. \cite{biswas} and that the fact that 
varying $a$ could improve the numerical efficiency was found empirically 
in Ref. \cite{kbeck81}. Eq. (\ref{eq:optim}) can be used to understand 
these results. For instance, for $H=p^2+x^2+x^8$, we can obtain a very 
accurate result with $x_{max}=2.8$ (see Section \ref{sec:app}).
According to Eq. (\ref{eq:optim}) the optimal value of $a$ in this case is
$a=(2/5)x_{max}^3\simeq 8.8$ which is slightly below the value $(\approx 10)$
suggested in \cite{kbeck81}. Note also that equivalently good results 
can be obtained using $G=-bx^3$. 
\section{Approximate error formulas}
\label{sec:app}
In this Section,we discuss the intrinsic error
$\delta E =E(x_{max})-E(\infty)$ 
where $E(x_{max})$ is defined by $\psi(x_{max},E(x_{max}))=0$,
for a given energy level. We emphasize that $\delta E$ is the error due to
the finiteness of $x_{max}$ independently of practical considerations 
regarding the numerical estimation of $E(x_{max})$ which is assumed to be 
known with an error much smaller than $\delta E$ in this section.
We use the familiar parametrization of the
quadratic term of the potential, $V_2={1\over 2}m\omega^2$ and we restore the 
dependence on $\hbar$ and $m$.
The error for the ground state of the harmonic oscillator has been 
estimated in Eq. (4) of Ref. \cite{convpert}.
Using the asymptotic form of the integral in this equation, we obtain 
\begin{equation}
\delta E_0^{harm.} \simeq 2 \bigl({S_0\over{\pi \hbar}}\bigr)^{1\over2}{\rm e}^
{-{S_0/{\hbar}}}\ ,
\label{eq:harmerr}
\end{equation}
with 
\begin{equation}
S_0=\int_{-\infty}^{+\infty}dt{1\over 2}m((\dot{x}_c(t))^2+\omega^2(x_c(t))^2)
=m\omega x_{max}^2
\end{equation}
and $x_c(t)=x_{max}{\rm e}^{-\omega |t-t_0|}$.
This corresponds to semi-classical approximation where the contribution of 
the large field configurations
are obtained by calculating the quadratic fluctuations with respect 
to $x_c(t)$.
The anharmonic corrections can 
be approximated to lowest order in the the anharmonic couplings
by adding a term $S_{anh}$ to $S_0$ in the exponent of Eq. (\ref{eq:harmerr}) 
with
\begin{equation}
S_{anh}=\int_{-\infty}^{+\infty} dt V_{anh}(x_c(t))\ ,
\end{equation}
and $V_{anh}$ is the anharmonic part of the potential.
Our final perturbative estimate is thus
\begin{equation}
\delta E_0 \simeq \delta E_0^{harm.}
{\rm e}^{-\sum_{j=2}^{l}({1\over{j\hbar }})V_{2j} x_{max}^{2j}}
\label{eq:dele}
\end{equation}
This estimate is accurate if the $V_{2j}$ and $x_{max}$
are small enough. We expect that for the excited states, approximate formulas 
of the form of Eq. (\ref{eq:dele}) multiplied by a polynomial should hold.

When $\lambda$ or $x_{max}$ become too large, Eq. (\ref{eq:dele}) 
is not adequate. To obtain a better approximation, we use
\begin{equation}
{\partial\over{\partial x_{max}}}\psi(x_{max},E(x_{max}))=0\ ,
\end{equation}
and the asymptotic behavior of $\Psi$.
We estimate that $\partial \Psi/\partial E$ is of the 
order of the non-normalizable WKB solution and as a consequence, 
$\delta E$ 
has the asymptotic form
\begin{equation}
\delta E\simeq P(x_{max})(\psi(x_{max}))^2 \ ,
\end{equation}
where $P$ is a polynomial. This form is correct for the 
ground state of the harmonic
oscillator.
In the case where the leading term of $V$ is $V_{2l}x^{2l}$, 
this implies the asymptotic order of magnitude
estimate
\begin{equation}
\delta E\approx {\rm e}^{-{2\over{(l+1)\hbar}}\sqrt{2mV_{2l}}x_{max}^{l+1}}\ ,
\label{eq:cufit}
\end{equation}

\begin{figure}
\centerline{\psfig{figure=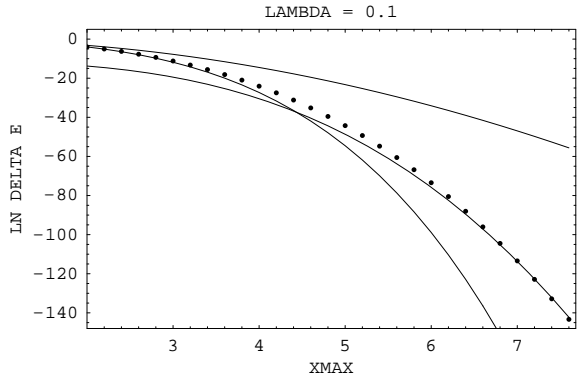,width=3.in}}
\caption{{\rm ln}( $\delta E_0$) as a function of $x_{max}$ 
for $\lambda$=0.1 (black dots). 
%(lower set of points). 
The continuous lines are 
from top to bottom on the left of the figure: 
the harmonic case (Eq. (\ref{eq:harmerr})), 
Eq. (\ref{eq:dele}) with $V_4=0.1$ (fits the dots well on the left of the 
figure), Eq. (\ref{eq:cufit}) (fits the dots well on the right of the 
figure).}
\label{fig:varx}
\end{figure}

We have tested the two approximate errors formulas given
above (Eqs. (\ref{eq:dele}) and (\ref{eq:cufit})) for 
the ground state corresponding to $V_{anh}=\lambda x^4$. 
We used the numerical values $\hbar=m=\omega=1$ and $\lambda=0.1$
The results are shown in
Fig. \ref{fig:varx}. 
We see that for small values of $x_{max}$, the perturbative estimate of 
Eq. (\ref{eq:dele}) corrects properly the harmonic result. 
However when  $x_{max}$ increases, the Eq. (\ref{eq:cufit}) gives better 
results. If the left part of the graph is displayed with a log-log scale,
it is approximately linear with a slope close to 3. In Fig. (\ref{fig:varx}),
the proportionality constant not given by Eq. (\ref{eq:cufit}) has been 
determined by fitting the 5 last data points on the right of the figure.
We conclude that by combining the two approximations it is possible to 
get a reasonable estimate of the errors on $E$ over a wide range of $x_{max}$.
We have tested Eq. (\ref{eq:cufit}) for other potentials. For instance,
for $H=p^2+x^2+x^8$, in order to get 30 significant digit, 
we estimated that $x_{max}\simeq 2.8$. We found that the difference 
between the ground state energy found from the condition $K=0$ (upper 
bound) and $L=0$ (lower bound) differed in the 30th significant digits.
\section{A challenging test}
\label{sec:chall}
The only practical limitation of the method proposed here is that 
in some cases the relevant details of the potential appear in widely separated
regions, forcing us to calculate a huge number of coefficients with many 
significant digits.
A simple example where such problem may occur is the symmetric double-well
with a small quartic coupling where the separation between the 
wells goes like the inverse square root of the quartic coupling.

In Ref. \cite{jentschura}, the lowest even and 
odd energies were calculated for a potential with $m=1,\ \hbar=1, 
\ V_2=-1/4,\ V_4=1/2000$ with 180 significant digits. Remarkably, the 
authors were able to reproduce the 110 significant digits of the splitting
between these two states 
by calculating instanton effects. 
We have reproduced the 180 digits of both states using an expansion
of order 1700 for $K$ and a value of $x_{max}=46$. The calculations were 
performed with 700 digit arithmetic. The calculation of one level with
such a procedure takes less than two hours
with MATHEMATICA on an unexpensive laptop using Pentium3. 
The computation time increases
with the accuracy required. In order to fix the ideas, 
it takes less than 2 minutes 
minutes to reproduce the 
first 120 digits in the above calculation.
\section{The multivariable case}
\label{sec:multi}

The basic equations presented in Section \ref{sec:basic} can be extended 
when the single variable $x$ is replaced by a $N$-dimensional vector $\vec{x}$.
In Eq. (\ref{eq:repa}), $\phi$ becomes a vector $\vec{ \phi}$ and the integral
a line integral. In order to guarantee that the wave function is independent
of the choice of the line, we require that the curl of $\vec{\phi}$ vanishes.
Eq. (\ref{eq:ric}) becomes:
\begin{equation}
\hbar \vec{\nabla} \vec{\phi} =\vec{\phi}\cdot\vec{\phi}+2m(E-V)\ .
\label{eq:ricmany}
\end{equation}
Using $\vec{ \phi}=\vec{ L}/K$, we write as previously 
\begin{eqnarray}
\hbar \vec{\nabla}\vec{ L}&+&2m(V-E)K+\vec{G}\cdot\vec{L}=0
\label{eq:v1} \\
\hbar \vec{\nabla}  K&+&\vec{L}+\vec{G}K=0 \ ,
\label{eq:v2}
\end{eqnarray}
with $\vec{G}(\vec{x})$ unspecified at this point.
These equations imply the multivariable Riccati equation (\ref{eq:ricmany})
multiplied by $K^2$. Near a zero of $K$, these equations imply the 
same singularity as Eq. (\ref{eq:ricmany}). 
After using Eq. (\ref{eq:v2}), the condition that
$\phi$ has no curl reads 
\begin{equation}
\nabla_i L_{(j)}+G_{(i)}L_{(j)}=\nabla_jL_{i}+G_{(j)}L_{(i)}\ .
\end{equation} 
The parenthesis for the vector indices are used in order to distinguish 
these indices from the order in a power series expansion used later.

The transformation Eqs. (\ref{eq:gt}) can vectorized trivially with $Q$ treated
as a scalar. In the expression of $K$ given by Eq. (\ref{eq:kgen}), 
the integral becomes a line integral and we require  
that $\vec{G}(\vec{x})$ has a vanishing curl. This condition is also necessary
to establish that different derivatives acting on $K$ commute.

The choice of coordinates to be used depends on the choice of boundary
conditions imposed. If we require $\Psi $ to vanish on a large 
hypersphere,  hyperspherical harmonics should be used. If we require $\Psi$ 
to vanish on hypercubes (as suggested for lattice problems in Ref. 
\cite{convpert}) cartesian coordinates should be used. To fix the ideas, 
let us consider the case of cartesian coordinates for two variables 
$x_1$ and $x_2$ with boundary conditions on a rectangle. 
We expand $K(x_1,x_2)=\sum_{m,n\geq 0}K_{m,n}x_1^mx_2^n$ 
and similar
expansions for the two components of $\vec{L}$. 
The coefficients can be constructed order by order, with the order of 
$K_{m,n}$ defined as $m+n$. The terms with one derivative yield the 
higher order terms. For instance, for $K$,
we obtain equations providing $K_{m+1,n}$ and $K_{m,n+1}$ in terms 
of coefficients of higher order just as in Eq. (\ref{eq:iter}). 
A detailed construction shows that if 
$V(x_1,x_2)$ has no special symmetry, we can determine all the  
coefficient up to a given order $l$ 
provided that we supply the values of two coefficients at
each intermediate order (for instance $\vec{L}_{m,0}$  for $m\leq l$).
These coefficients together with $E$ are fixed by the boundary 
conditions $K(x_{1min},x_2)=K(x_{1max},x_2)=K(x_1,x_{2min})=K(x_1,x_{2max})=0$.
Taking derivatives with respect to the free variables $x_1$ and $x_2$, 
and setting these variables to 0, we
obtain an infinite set of conditions. The truncation of this set, 
together with the truncation of the expansion in the other variable 
must be studied carefully. If we consider the special case where the 
problem can be solved by separation of variables, we  see that it is
important to maintain a uniform accuracy for all the conditions. 
If all the coefficients have been calculated up to order $l$, this can 
achieved in the following way.
We retain of the order of $l/2$ derivatives of the four conditions in such
a way that we get exactly $2l+3$ conditions which can be expanded up to 
an order close to  
$l/2$ in the remaining variable. A practical implementation of this program
is in progress.

\section{Conclusions}

In conclusion, we have shown that accurate estimates of the energy levels 
of arbitrary polynomial potentials  
bounded from below can be obtained by solving
polynomial equations. The fact that the function $L$ and $K$ are entire
guarantees that if we calculate enough terms we will gain proper control
of the asymptotic behavior of the wave function. Reaching this goal is 
in general a difficult task which often requires guesswork and analytical
continuations (see e.g., Ref. \cite{bender96}). Here, the convergence of
the procedure is guaranteed and the order at which we can terminate the 
expansion in order to reach a given accuracy can be estimated.
In addition, a systematic understanding 
and control of the errors due to the finite value of $x_{max}$ has been 
achieved.

The understanding of the gauge invariance of the basic equations proposed
here completely resolves the issues raised from our initial proposal 
\cite{bacus}. By varying $G$, from 0 to -$\phi$, we can interpolate
between a situation where $K$ is the wave function to another situation 
where $K=1$ and $L =\phi$. However, for every other choice of $G$, only 
the ratio $L/K$ has a direct physical meaning. 
By properly chosing $G$, we can at the same time improve the convergence
of $K$ and amplify the bifurcation toward the the non-normalizable 
behavior.

The extreme accuracy obtained for two widely separated wells indicates that 
for reasonably complicated potential, the number of terms that
needs to be calculated is not prohibitive. We intend to use this method
to test analytical results regarding the role of large configurations 
in the path-integral and to test semi-classical treatment of potentials 
with asymmetric wells \cite{coleman,zj}.

The method can be extended in the case of several variables. It remains to be 
determined if the simultaneous solution of many polynomial equations 
can be accomplished with a reasonable accuracy. For these problems, the 
fact that a judicious choice of the arbitrary functions $\vec{G}$ allows
to decrease the order of the expansions may be crucial.

\begin{acknowledgments}
We thank B. Oktay for communicating his work regarding the treatment of 
parity non invariant potentials with the method of Ref. \cite{bacus}.
We thank P. Kleiber, W. Klink, L. Li,
G. Payne, W. Polyzou, M.H. Reno and V.G.J. Rodgers for valuable conversations.
We thank F. Fernandez for pointing out Ref. \cite{kbeck81} to us.
This research was supported in part by the Department of Energy 
under Contract No. FG02-91ER40664,
and in part by the CIFRE of the University of Iowa.

\end{acknowledgments}
%\bibliography{da} 
%\bibliographystyle{unsrt}
%\bibliographystyle{prsty}
%\end{multicols}
%\end{document}

\end{multicols}
\end{document}